\documentclass[twocolumn,prl]{revtex4}
\usepackage{graphicx}
\usepackage{dcolumn}
\usepackage{bm}
\usepackage{amsmath}

\begin{document}

\title{Optical RKKY Interaction between Charged Semiconductor Quantum Dots}

\author{C. Piermarocchi} \author{Pochung Chen} \author{L. J. Sham}
\affiliation{Department of Physics, University of California San
Diego, La Jolla, California 92093-0319} \author{D. G. Steel}
\affiliation{Harrison M. Randall Laboratory of Physics, The University
of Michigan, Ann Arbor, MI 48109-1120}

\date{\today}

\begin{abstract} 
We show how a spin interaction between electrons localized in
neighboring quantum dots can be induced and controlled optically. The
coupling is generated via virtual excitation of delocalized excitons
and provides an efficient coherent control of the spins.  This quantum
manipulation can be realized in the adiabatic limit and is robust
against decoherence by spontaneous emission. Applications to the
realization of quantum gates, scalable quantum computers, and to the
control of magnetization in an array of charged dots are proposed.

\end{abstract}

\maketitle

Quantum control of an electron spin, either independent of other spins
or condition on their states, in a semiconductor nanostructure is a
central issue in the emerging fields of spintronics and quantum
information processing.  The spin of a single electron confined in a
semiconductor quantum dot (QD) was proposed \cite{loss98} as a qubit
for the realization of scalable quantum computers. Quantum gates are
designed using electric gates to control via overlap the exchange
interaction between two electrons in neighboring dots.  Optical
control was also proposed, in which a cavity mode couples different
dots \cite{imamoglu99}, or a dipole-dipole interaction between charged
excitons strongly polarized by an external dc field is exploited \cite
{pazy01}.  Optical control possesses several advantages compared with
control by gate voltage. Ultrafast lasers can control quantum systems
on the femtosecond time scale, and using shaping techniques the
amplitude and phase of the pulses can be designed at will offering a
great deal of flexibility and efficiency \cite{chen01}.

In this paper we report a theory of an exchange interaction between
two electron spins in separate dots in a typical semiconductor QD
system by virtual excitation of delocalized exciton states in the
host material which interact with the electrons in both dots.  This
time-dependent effective interaction is driven by the external laser
field, and is, thus, controllable.  The virtual excitation by
an off-resonant laser preserve the coherence of the spin dynamics.
This indirect exchange mechanism is analogous to a RKKY interaction
\cite{rudemann54} between two magnetic impurities mediated by
conduction electron or excitons\cite{melo95}, except that the
intermediate electron-hole pair is produced by the external light.
The optical quantum control of a single exciton in a semiconductor QD
has been recently reported in GaAs QDs generated by monolayer
fluctuations \cite{stievater01} and InGaAs self-assembled QDs
\cite{kamada01,htoon02}. The short radiative recombination lifetime of
the exciton (of the order of 100~ps) gives a severe limitation for the
application to quantum computation, even with the help of shaping
techniques \cite{chen01}. This can be avoided by doping QDs each with
a single conduction electron and by encoding the quantum information
in the spin degrees of freedom. Optical control by virtual excitation
avoids the fast optical decoherence.  Thus, the advantages of a very
long spin coherence time in QDs \cite{gupta99} and fast optical
control can be combined.

Consider two electrons localized in two QDs at $\mathbf{R}_\ell$
($\ell=1,2$) with wavefunctions
$\phi_\ell(\mathbf{r}-\mathbf{R}_\ell)$ and a laser field used to
generate exciton states in a continuum. This continuum is provided by
states in the host material embedding the QDs with an energy gap
$\epsilon_{\text{G}}$. QDs can be embedded in bulk, quantum well, or
quantum wire host structures.  Fluctuation QDs embedded in a narrow
quantum well \cite{stievater01} represent an example of a system with a
two dimensional continuum.  A promising system for the scheme which we
propose is provided by pyramidal QDs \cite{hartmann00}, where
localized states in growth-controlled QDs and delocalized states in
the so-called vertical quantum wire are well separated and can be
addressed selectively. In this case the continuum states are in the
vertical quantum wire.  The Coulomb interaction between the
photoexcited pairs and the localized states contains direct and
exchange contributions. The direct term gives state renormalization.
The attraction of the exciton to the dot is determined in the long range by
the dipole moment of the exciton induced by the localized electron in the dot
and in the short range by the dot potential. The binding of the exciton to
the dot is then sensitive to the design of the dot, ranging from a very
weakly bound state to one localized in the dot \cite{chen03}. The former has
a wave function overlap to the neighboring dot and contributes to the
optical RKKY. The latter can be made far off resonance to the optical
excitation. Here we focus on the spin structure of the
Hamiltonian arising out of the exchange interaction between the
localized and optically excited conduction electrons.  The exchange
interaction between the localized electron and the valence hole is
negligible. For convenience, $\hbar=1$ throughout.  Hence,
the Hamiltonian of the system contains, besides the electron and hole
energies in the host and in the dots, (1) the exchange interaction between
electrons,
\begin{equation}
H_{\text{X}}=-\frac{1}{V} \sum_{\ell,\alpha,\alpha^\prime
\atop{\mathbf{k},\mathbf{k}^\prime}}
J_\ell(\mathbf{k},\mathbf{k}^\prime)
~\mathbf{S}^\ell\cdot\mathbf{s}_{\alpha,\alpha^\prime}
~c^{\dagger}_{\mathbf{k},\alpha} c_{\mathbf{k}^\prime,\alpha^\prime},
\label{HX}
\end{equation}
where $\mathbf{S}^j$ denotes the spin of the $j$-th localized electron,
$\mathbf{s}$ the spin of the electron in the photogenerated pair and
$c^{\dagger}_{\mathbf{k},\alpha}$ and
$h^{\dagger}_{\mathbf{k},\beta}$ are the creation operators of free
electron and hole spin states, respectively; and
(2) the time-dependent control Hamiltonian describing the creation of
electron-hole pairs 
\begin{eqnarray}
H_{\text{C}}(t)=\sum_{\mathbf{k},\sigma} \frac{\Omega_{\mathbf{k},\sigma}(t)}{2}
e^{-i\omega_{P\sigma}t}
c^\dagger_{\mathbf{k},-\sigma} h^\dagger_{-\mathbf{k},\sigma}
+ \text{h.c.}~.
\end{eqnarray}
$\Omega_{\mathbf{k},\sigma}(t)$ is the time-dependent Rabi energy
associated with the electric field of the optical pulse times the
transition dipole matrix element of the electron-hole pair with momenta $\pm
\mathbf{k}$, resulting from taking the wave vector of the photon to be
zero.  $\sigma=\pm$ denotes the $\sigma\pm$ circular polarization of light,
which fixes the spin configuration of the photoexcited electron spin state
$-\sigma(1/2)$ and hole state $\sigma (3/2)$.
We consider only a single heavy hole band  which is valid for GaAs
confined heterostructure.  
For the exchange integral $J_\ell(\mathbf{k},\mathbf{k}^\prime)$, the 
Coulomb interaction is screened by the static dielectric constant
\cite{sham66}.  We approximate the continuum electron by a
 plane wave orthogonalized  (OPW) to the dot states and further
simplify the exchange to the form
\begin{eqnarray}
J_\ell(\mathbf{k},\mathbf{k}^\prime) &\equiv &
j_\ell e^{-i(\mathbf{k}-\mathbf{k}^\prime)\cdot\mathbf{R}_\ell},
\end{eqnarray}
with a constant prefactor given by
\begin{equation}
j^{\text{d}}_\ell \sim I Ry^* a_{\text{B}}^* \xi^{d-1},
\label{estj}
\end{equation}
where $\xi$ is the localization length within a dot, $Ry^*$ and
$a_{\text{B}}^*$ denote the effective exciton Rydberg energy and Bohr radius
in the host semiconductor, $I$ is a dimensionless constant that depends only
on the particular geometry of the dot, and
$d$ is the dimensionality of the host. The dependence of $I$ on the wave
vectors is removed by using a suitable average discussed below.

\begin{figure}[t]
\parbox{70 mm}{\includegraphics{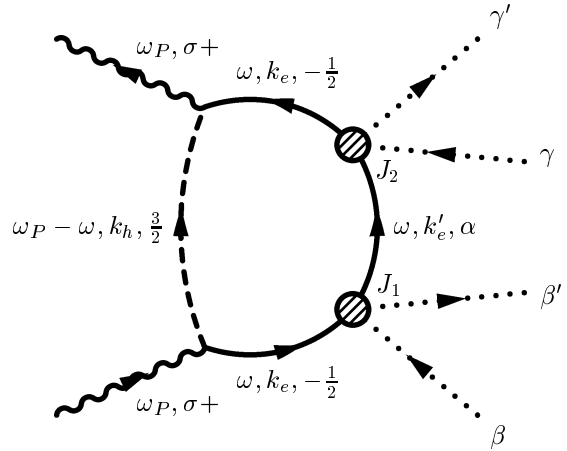}}
\caption{Effective spin-spin interaction for the localized electrons
in the dot 1 and 2 (indicated by dotted lines) induced by a
photoexcited electron-hole pair (the solid and dashed line,
respectively).  The indices $\beta$ and $\gamma$ denote the spin
states of the electrons localized in the dots. The photon propagator
is depicted by a wavy line.}
\label{fig.1}
\end{figure}

In the absence of the laser pulse, the system with two localized
electrons is in a degenerate ground state with four spin states. In the 
presence of a laser field nearly resonant with the continuum, the
ground state energy is shifted by $\Delta E_0= - \sum_{\mathbf{k}}
(\Omega_{\sigma}(t)/2) (\delta+
k^2/2\mu)^{-1}(\Omega^*_{\sigma}(t)/2)$ where
$\delta=\epsilon_{\text{G}}-\omega_{\text{P}}$ is the detuning of the
laser with respect to the electron hole continuum and $\mu$ is the
reduced mass of the electron hole pair. $\Delta E_0$ contributes to
the well-known blue shift in the excitonic transition or dynamic Stark
effect. This contribution, which is diagonal in the spin index of the
localized states, is irrelevant for our spin control
purposes. However, when Coulomb interaction is taken into account, the
spin configuration of the localized electrons does affect this 
shift. To calculate this effect we break the pair propagator $(\delta+
k^2/2\mu)^{-1}$ into its electron and hole parts, and then consider
the self-energy correction in the electron propagator due to the
interaction with the localized states. These corrections give an
effective Hamiltonian for the spins of the localized electrons.
Fig.~\ref{fig.1} shows the lowest order contribution in Coulomb
interaction to the effective exchange interaction between the two
localized electrons mediated by the free electron in the photoexcited
pair.  The incoming photon (wavy line) with energy $\omega_{\text{P}}$
and polarization $\sigma+$ creates a pair of electron (solid line) and
hole (dashed line). The electron in the pair interacts with the two
electrons localized in the two neighboring dots (dotted lines) via
$H_X$ in Eq.~(\ref{HX}), and then recombines with the hole. The
indices $\beta$ and $\gamma$ refers to the spin states of the
localized electrons.  The loop contains the integral in the exchanged
energy $\omega$, and the sum in the momentum $\mathbf{k}^\prime_e$ and
spin $\alpha$ of the intermediate electron. If the free carriers'
motion is spin independent, the interaction between two local spins
associated with this diagram contains the term
$(\mathbf{S}^1\cdot\mathbf{s}) (\mathbf{S}^2\cdot\mathbf{s})$ $=
(\mathbf{S}^1\cdot\mathbf{S}^2)/4 +i\mathbf{s}\cdot(\mathbf{S}^1\times
\mathbf{S}^2)/2.$ When it is summed with the diagram in which the
order of the two local spins is reversed, the cross product terms
cancel.  Hence, the effective spin-spin Hamiltonian assumes the
Heisenberg form
\begin{equation}
H_s = - 2 J_{12} \, \mathbf{S}^1\cdot\mathbf{S}^2,
\label{heff}
\end{equation}
where the effective exchange constant $J_{12}$ is always positive, given by
\begin{widetext}
\begin{equation}
J_{12}(R)= \frac{1}{16} \sum_\sigma |\Omega_{\sigma}(t)|^2  j^{\text{d}}_1
j^{\text{d}}_2
 \int\frac{d^d\mathbf{k}}{(2\pi)^d} \frac{d^d\mathbf{k}^\prime}{(2\pi)^d}
 e^{-i(\mathbf{k}-\mathbf{k}^\prime)\cdot\mathbf{R}} \left(\delta +
 \frac{k^2}{2m_{\text{h}}} + \frac{k^2}{2m_{\text{e}}} \right)^{-2}
 \left(\delta + \frac{k^2}{2m_{\text{h}}} + \frac{k^{\prime
 2}}{2m_{\text{e}}} \right)^{-1} ,
\label{spatial}
\end{equation}
\end{widetext}
where  $m_{\text{e}}$ ($m_{\text{h}}$) denotes the electron (hole) mass, and
$\mathbf{R}=\mathbf{R}_{1}-\mathbf{R}_{2}$. The Rabi energy can
reasonably be assumed independent of $\mathbf{k}$.  From the
Eq.~(\ref{spatial}) it can be seen that $J_{21}=J_{12}$.  Parabolic
dispersions for the electron and the hole energies are assumed. 
Unpolarized light is used to induce the ferromagnetic interaction
between the local spins, so as to avoid the spin polarization effect of the
first order process in $H_X$.

The most important correction to Fig.~\ref{fig.1} is due to the
Coulomb attraction between the electron and hole in the excited
pair \cite{melo95}, giving rise to three exciton propagators.  The
electron-hole pair energies in Eq.~(\ref{spatial}) are replaced by the
exciton energies. The exciton wave functions enhance the oscillator
strength and, hence, the Rabi energies at the two optical vertices and
the exchange constants at the spin vertices.  To remain in the regime
of virtual excitations the laser energy must be adjusted to below the
lowest discrete exciton state. Keeping only the $1s$ exciton
contribution we obtain
\begin{equation}
J_{12}=\left(\frac{\Omega}{4 \delta}\right)^2 \frac{j^{\text{d}}_1
j^{\text{d}}_2 |\phi^{\text{d}}_{\text{1s}}(0)|^2}{\delta} I_{\text{d}}(R),
\label{J12}
\end{equation}
\begin{equation}
\mbox{where~~}I_{\text{d}}(R)=
\int
\frac{d^d\mathbf{q}}{(2\pi)^d} 
\frac{e^{i\mathbf{q}\cdot{R}}[\rho_{\text{1s}}^{\text{d}}(\alpha 
q)]^2}{1+(\kappa_{\text{M}} q)^2}.
\label{integral}
\end{equation}
The term $\phi^{\text{d}}_{\text{1s}}(0)$ is the 1s excitonic
wavefunction in $d$ dimensions, $\kappa_{\text{M}}=1/\sqrt{ 2 M
\delta}$ is a characteristic optical length related to the detuning,
$M$ is the total mass of the pair, and $\alpha=m_{\text{h}}/M$.  The
integral in $\mathbf{q}$ represents the sum over the exciton center of
mass wavevectors scattered by the localized electrons. The form factor
is defined as $\rho_{\text{1s}}^{\text{d}}(q)=\int d^d \mathbf{r}
|\phi^{\text{d}}_{\text{1s}}(\mathbf{r})|^2
e^{i\mathbf{q}\cdot\mathbf{r}}~.$ The dependence of the interaction on
the interdot separation is contained in the integrals
$I^{\text{d}}(R)$. Their explicit analytical expression and the
contributions from higher excitonic levels will be given in a long
publication \cite{chen03}. The spatial dependence
in all three cases, $d=1,2,3$, is dominated by two exponential terms
with two characteristic lengths: the optical length $\kappa_{\text{M}}$ and
the Bohr radius.  

\begin{figure}[t] \includegraphics[scale=0.5]{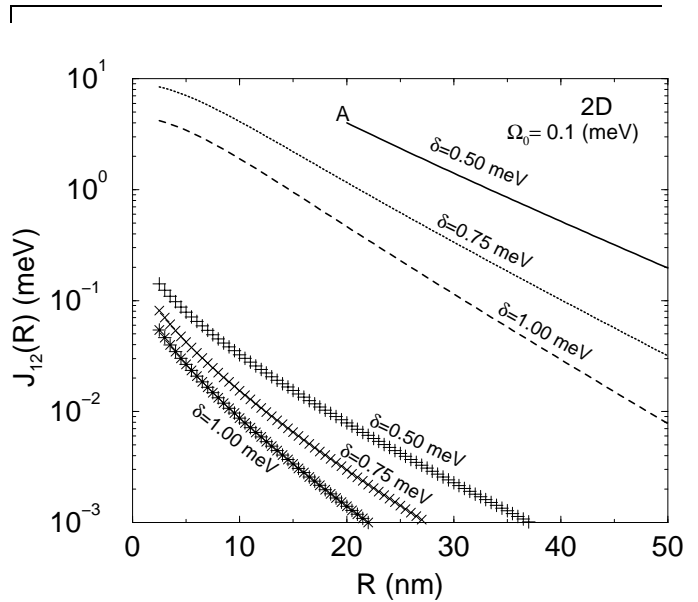}
\caption{Optically induced interaction in the two dimensional host as
a function of the distance between the dots, with the excitonic
corrections (upper 3 lines) and without (lower 3 curves), for three
detuning values.  The upper limit, if reached, of the adiabaticity
condition in Eq.~(\ref{adiab}) for a swap gate is indicated by Label
A. $\Omega_0$ represents the peak value of the Rabi energy in the
pulse.}  \label{fig2} \end{figure}

In Fig.~\ref{fig2} we plot $J_{12}$ as a function of the separation
between the dots in the two dimensional case, for three different
values of the detuning. The results when excitonic effects are not
included are also shown for comparison.  The excitonic enhancement can
be  more than two order of magnitudes. An analogous excitonic
enhancement effects was found in the case of spin-flip Raman
scattering of electrons trapped in neutral donors \cite{thomas68}.  In
the calculation we used $m_e=0.07 m$ and $m_h=0.5 m$, a Rabi energy of
0.1 meV at the peak of the pulse, 5 meV for the exciton Rydberg, and
the dot dimension $ \xi \sim 2 a_B^*$.  A typical value for the Bohr
radius is 150 \AA.  A simple estimate of $I=5.76$, the magnitude of
the exchange in Eq.~(\ref{estj}), is obtained for one OPW from the
average of exchange $j_\ell$ squared over $k$ and $k'$ for the wave
vector cutoff equal to $1.1/\xi$. This value corresponds to the
minimum value of the averages for all possible cutoffs. An orbital of
Slater form is assumed for the localized electron. The exchange energy
between two local spins is then of the order of 1~meV, which is
comparable to the estimated values of the exchange coupling due to
tunneling in coupled QDs \cite{burkard99}.  The interaction can be
considerably enhanced in systems where large QDs are vertically
stacked with separations smaller that their lateral size.
 
The detuning $\delta$ has to be larger than the exciton line-width so
that there is no absorption of energy in the exciton spectrum from the
pump. The dynamics of the localized spin is thus fully coherent and
the finite lifetime $\Gamma$ of the photoexcited states is no longer a
limitation for quantum information processing.  The optical pulses has
to be switched on and off in such a way that no real population is
excited in the intermediate excitonic states. In the simple case of a
two-level system $G$ and $E$ this adiabaticity condition can be
written in the form$ |(\frac{d}{dt}\langle G^\prime|)|E^\prime
\rangle| \ll |\epsilon_{G^\prime}-\epsilon_{E^\prime}|$ where
$|G^\prime \rangle$ and $|E^\prime \rangle$ are the time dependent
dressed states in the rotating frame (or time-dependent Floquet
states) of the ground and excited state \cite{bayfield99}. This leads
to $ |\delta~ \dot{\Omega}(t)|\ll (\delta^2+\Omega(t)^2)^{3/2}~.$
Assuming Gaussian optical pulses of the form $\Omega(t)=\Omega_0
e^{-(t/\sigma)^2}$, we find the condition for the length of the pulse
to preserve the adiabaticity to be \begin{equation} \sigma \gg
\frac{\Omega_0}{\delta^2} .  \label{adiab} \end{equation} If we fix
the angle of rotation of the spin by, e.g., $\int~dt J(t)=\pi/2$ (swap
gate \cite{loss98}), the condition above can be translated into an
upper limit for $J_{12}$. This limit is indicated explicitly in Fig. 2
by the label ``A'' in the top curve. The other curves shown are all
within the adiabatic limit.  Our case differs from the ideal two-level
case discussed above in two aspects: the presence of a continuum of
excitonic states in addition to the discrete states and the
interaction of the continuum with the localized states.  Neither
changes considerably the estimates which go into the adiabaticity
condition and Eq.~(\ref{adiab}) remains valid. The exchange Coulomb
interaction between the photoexcited electron and the localized
electrons in Eq.~(\ref{HX}) is time independent and does not affect
the adiabaticity.  For the continuum, we have used a Fano approach and
defined $ |G^\prime\rangle$ and $|E^\prime,\mathbf{k}\rangle$ as the
ground state and continuum states dressed by the external laser
field. Because of the detuning and the neglect of higher-order
many-body effects, the Fano problem can be solved exactly.  The
dressed states can be explicitly written as a function of the density
of states of the continuum and the coupling. The adiabaticity
condition can then be put in the form $(\frac{d}{dt}\langle
G^\prime|)|E^\prime,\mathbf{k}\rangle \ll |\delta|;\forall~ \mathbf{k}
~$.  It is possible to prove \cite{chen03} that, for a reasonable
density of states and energy dependence of the optical coupling, it is
sufficient to satisfy this condition for $\mathbf{k}=0$. This reduces
to a two-level problem where Eq.~(\ref{adiab}) holds as a sufficient
condition for adiabaticity.

A simple experimental setup with two dots of different sizes and a
tunable laser could be used to check the spin entanglement from the
interaction proposed here. The manipulation and measurements on a
single dot can be realized selectively by exciting at the energy of
the localized exciton.  Single qubit operation on a selected dot can
be carried out in a Raman configuration as proposed for instance in
Ref. \cite{imamoglu99}. An external magnetic field has to be used to
initialize the system.  Quantum computation requires controllable
spin-spin interaction between  specific nearest neighbor pairs.  Our
proposed mechanism can induce spin-spin interaction.  Near-field
optical excitations can be used to ensure the interaction is between
nearest neighbor.  The dots in fact can be arranged in an array
separated by distances of the order of the wavelength of light.  Then,
quasi-delocalized exciton states with sub-wavelength extension can be
excited only between the two dots intended to be coupled. This
possibility is well within the experimental state-of-the-art
capabilities of near-field scanning optical microscopy \cite{guest01}.
In such a system, universal quantum computation can in principle be 
realized by using only the optically controlled exchange interaction,
without resorting to single qubit operation \cite{divincenzo00}.  As a
possible extension to spintronics, we suggest that a regular array of
charged dots may be magnetized by light which initializes the
magnetization by exciting conduction electrons.

In conclusion we have proposed an optical technique to generate and
control the entanglement of the spin of two electrons localized in
neighboring QDs. The control of the spin can be realized in the
adiabatic regime and can lead to the realization of spin quantum gates
useful for quantum information processing and to control of the
magnetization of an ensemble of dots.

\acknowledgments This work was supported by ARO F0005010, NSF
DMR-0099572, and DARPA/ONR N0014-99-1-109. We thank Dr. D. Gammon for
stimulating discussions.

\end{document}